\documentclass[aps,prl,reprint]{revtex4-1}

\usepackage{graphicx} 
\usepackage[english]{babel}
\usepackage[T1]{fontenc}
\usepackage[usenames,dvipsnames]{xcolor}
\usepackage{setspace}
\usepackage[compact]{titlesec}
\usepackage{hyperref}
\usepackage{tikz}
\usepackage{amsmath}
\usepackage{amssymb}
\usepackage{mathtools}
\usepackage{appendix}
\usepackage{lipsum}
\usepackage{}
\begin{document}

\title{Entropic Colloidal Crystal Prediction: A Quantum Density Functional Theory Inspired Approach}

\author{Kristi Pepa$^{1,\dag}$, Isaac R. Spivack$^{2,\dag}$, Trevor F.G. Teague$^{1}$, Ryn Y. Oliphant$^{3}$, Domagoj Fijan$^{1}$,  and Sharon C. Glotzer$^{1,2,3,4}$}%
\email{sglotzer@umich.edu}

\affiliation{$^1$Department of Chemical Engineering, University of Michigan, Ann Arbor, Michigan 48109, USA.}
\affiliation{$^2$Department of Physics, University of Michigan, Ann Arbor, Michigan 48109, USA.}
\affiliation{$^3$Department of Materials Science \& Engineering, University of Michigan, Ann Arbor, Michigan 48109, USA.}
\affiliation{$^4$Biointerfaces Institute, University of Michigan, Ann Arbor, Michigan 48109, USA.}
\affiliation{$^\dag$K. Pepa and I. Spivack contributed equally to this work.  \\}


\date{\today}

\begin{abstract}
\normalsize
In pursuit of a colloidal analogue to quantum density functional theory (DFT) predictions of atomic crystal structures, we report a new, classical DFT that predicts the relative thermodynamic stability of colloidal crystals of hard, convex particle shapes. In contrast to standard classical DFT approaches, our theory maps the hard particle system to an auxiliary system in which we treat the particles as fixed ``nuclei'' embedded in a fictitious, spatially varying density field that distributes throughout the auxiliary system. By minimizing the free energy of the auxiliary system, and through comparison with known equations of state and free energy calculations using thermodynamic integration, we show that the auxiliary system with the lowest free energy corresponds to the most probable crystal of hard shapes in the original system.
\end{abstract}

\maketitle

Entropic forces alone can order particles into colloidal crystals of remarkable complexity and diversity. Hard-particle colloidal crystals isostructural to atomic and molecular crystals, including quasicrystals~\cite{haji-akbariPhaseDiagramHard2011,jeEntropicFormationThermodynamically2021}, clathrates~\cite{leeEntropyCompartmentalizationStabilizes2023}, cubic diamond crystals~\cite{damascenoCrystallineAssembliesDensest2012}, Frank-Kasper structures~\cite{damascenoPredictiveSelfAssemblyPolyhedra2012}, host-guest crystals~\cite{c.mooreShapedrivenEntropicSelfassembly2021}, crystals with as many as 432 particles in the unit cell~\cite{leeEntropicColloidalCrystallization2019}, and many more structures~\cite{dijkstraPhaseDiagramHighly1999, avendanoPackingEntropicPatchiness2017, agarwalMesophaseBehaviourPolyhedral2011}, have all self-assembled in molecular simulations -- and many in experiments~\cite{shevchenkoStructuralDiversityBinary2006, bolesSelfAssemblyColloidalNanocrystals2016,wangControlledSelfAssemblyGold2023} -- due solely to entropy maximization. The emergence of ordered structures from disordered fluids comprised solely of particles with no explicit interactions other than excluded volume is counterintuitive. Nevertheless, increased order can lead to increased entropy -- and thus decreased free energy -- when there are more microstates associated with the ordered macrostate than with the disordered macrostate at the same density~\cite{asakuraInteractionTwoBodies1954,onsagerEffectsShapeInteraction1949, frenkelUnderstandingMolecularSimulation2002,frenkelPerspectiveEffectShape2000,frenkelOrderEntropy2015}. Just as electronic interactions dictate the structure of atomic and molecular crystals, particle shape -- through entropy -- dictates the structure of colloidal crystals of hard particles. In the latter, entropy maximization results in the emergence of statistical, effectively attractive entropic forces among particles that act locally and directionally~\cite{vanandersUnderstandingShapeEntropy2014a}, much like a bond~\cite{harperEntropicBondColloidal2019}.

That identical structures are possible from very different forces and at very different length scales underscores the universal nature of statistical thermodynamics: ergodic systems evolve toward free energy minima regardless of the origin of the interactions. This fundamental tenet suggests the tantalizing possibility that understanding how and why atoms form the crystal structures they do can inform our understanding of why colloidal particles form the same crystal structures, and vice versa. Such insights could be game-changing in terms of materials design and prediction of both atomic and colloidal systems. This endeavor would be greatly facilitated if each could be described within comparable theoretical frameworks. Quantum density functional theory (qDFT) is one of the major workhorses for calculating and understanding the atomic and electronic structure of crystalline materials \cite{10.1119/1.19375, RevModPhys.87.897}. A comparable theory for colloidal crystals would be a classical density functional theory (cDFT). This motivates us to develop a classical DFT of entropic bonding in colloidal crystals that, through parallel construction with quantum DFT of electronic (chemical) bonding of atomic and molecular crystals, can provide an apples-to-apples comparison between hard particle colloidal crystals and their atomic and molecular counterparts. 

Over the years, many different cDFT treatments of hard particle systems have been proposed~\cite{PhysRevLett.63.980, Roth_2010, Hansen-Goos_2006, PhysRevE.102.062137, PhysRevE.102.062136, PhysRevLett.110.137801, Wittmann_2016}. In those theories, the particles are treated as comprising a structured solvent whose spatial dependence in the presence of an external perturbation such as a solute or interface is solved using one or another flavor of cDFT \cite{rothFundamentalMeasureTheory2010, 10.1063/1.1520530, hansen-goosDensityFunctionalTheory2006}. Those theories have been successful in the study of fluids and liquid crystals, and predict the liquid-solid phase transition of hard spheres \cite{PhysRevLett.63.980, Wittmann_2016, rothFundamentalMeasureTheory2010}. However, they have achieved limited success in predicting crystal phases of hard particles of non-spherical shape, largely because the approximation of solvent-solvent contributions to the free energy performs poorly at intermediate and high densities where crystals form~\cite{PhysRevE.102.062136, rothFundamentalMeasureTheory2010}. 

In this Letter, we present a new theory that treats the particles not as solvent, but as fixed solute particles surrounded by a fictitious density field that holds them in place. In this way, our formulation is closer in spirit to that of quantum DFT in which atomic nuclei comprise the fixed solute and electrons form a density field that hold the nuclei in place. In quantum DFT, the ground state electron density of an $N$-electron system of $M$ atoms is identical to that of an \textit{auxiliary} system of $N$ non-interacting electrons in an effective potential.  That effective potential contains an external potential term arising from the (assumed fixed) positively charged nuclei as well as terms describing electronic interactions, and must be  judiciously defined~\cite{kohnSelfConsistentEquationsIncluding1965}. Likewise, we posit that the thermodynamically preferred crystalline arrangement of a system of $M$ hard particles will be identical to that of an \textit{auxiliary} system comprised of a fictitious density field in an effective potential $V_{\text{eff}}$.  We further posit that the only contribution to $V_{\text{eff}}$ is an external field $V_{\text{ext}}$ arising from the positions and orientations of the $M$ fixed hard shape ``nuclei''.  

Our approach is as follows. We propose a new classical free energy functional for such an auxiliary system, assume an expression for $V_{\text{ext}}$, and simplify the resulting free energy functional. We then minimize the free energy to obtain the thermodynamically preferred spatial distribution of the fictitious density field in the auxiliary system for a set of candidate crystal structures. We hypothesize that embedded in this field is the information to predict the thermodynamically preferred crystal structure of the original hard shape system. We validate our theoretical predictions against known equations of state and/or Frenkel-Ladd free energy calculations for several example systems. To emphasize the very different nature of the classical density functional theory for hard particles derived here, we refer to our theory as EB-DFT for ``entropic bond'' DFT.
\begin{figure}[!ht]
\includegraphics*[width = 0.4\textwidth]{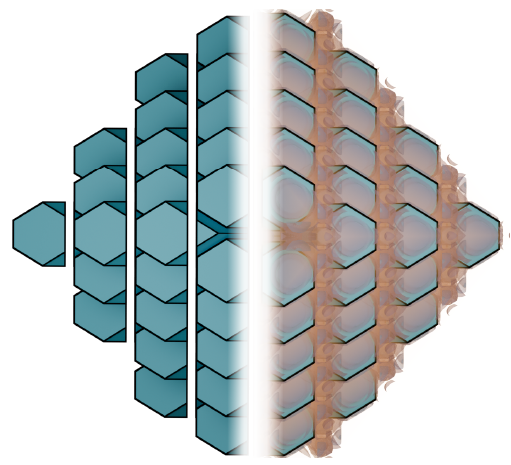}
   \caption{In an EB-DFT calculation, the original system (left) of hard particles (e.g.,  truncated tetrahedra assembled into a  cubic diamond crystal structure) is mapped to an auxiliary system (right) in which the hard particles are fixed in place and surrounded by a fictitious, spatially varying density field. The density field in the auxiliary system is optimized to minimize the EB-DFT free energy functional. We assert that the arrangement of hard particles that corresponds to the minimum energy of the auxiliary system corresponds to the maximum entropy of the original system.}
   \label{fig:sphere}
\end{figure}

\textbf{\textit{Derivation of EB-DFT --}}
We start with an auxiliary system in which a configuration of hard shapes is arranged and fixed in a proposed crystal macrostate. We imagine that all of the space in the auxiliary system not occupied by a fixed hard shape is filled with a fictitious density field, $\rho(\textbf{r})$, whose spatial dependence is dictated by a potential $V_{\text{ext}}$ arising from the positions and orientations of the fixed shapes. We posit that the fictitious density field be ideal, so that $V_{\text{ext}}$ is the only term we need to consider in our effective potential. We can then write the free energy of the auxiliary system as a functional of $\rho(\textbf{r})$ in the canonical ensemble~\cite{lutskoClassicalDensityFunctional2022}:
\begin{equation}
\begin{split}
    F[\rho(\textbf{r})] &= \int d\textbf{r} \rho(\textbf{r})\ln[\rho(\textbf{r})] -  \ln \left[ \frac{N^N}{N!}\right] \\ 
    &+  \int d\textbf{r} \rho(\textbf{r})V_{\text{ext}} + \lambda\int d\textbf{r} \rho(\textbf{r}) - \lambda N.
\label{Eq:pP_F}
\end{split}
\end{equation}
\noindent In Eq.~\ref{Eq:pP_F}, the first two terms comprise the entropic contribution to the free energy of the auxiliary system from an ideal N-particle density field. The last two terms in Eq.~\ref{Eq:pP_F} sum to zero and constrain the system to fixed $N$ (these terms would be absent in an ensemble where $N$ is not fixed). The third term accounts for the density-dependent interaction of $V_{\text{ext}}$ with the fictitious density field.

\textit{Form of the external potential-- }
In the spirit of Kohn-Sham DFT, we attempt to construct an expression for $V_{\text{ext}}$ that correctly maps the auxiliary system back to the original hard shape system. We imagine $V_{\text{ext}}$ to arise from the fixed hard shapes, which in the auxiliary system serve as boundary conditions. We first consider the distribution of an ideal density field in an arbitrary potential V(\textbf{r}):
\begin{subequations}
\label{Eq:V_ext}
\begin{align}
    \beta V(\textbf{r}) &= [\ln(\rho(\textbf{r})) - \beta \mu],
\end{align}
where $\mu$ is the chemical potential.
\end{subequations}
We conjecture that $V_{\text{ext}}$ can be expressed by modifying the above to include a spatially-dependent multiplicative factor, $\alpha(\textbf{r})$:
\begin{subequations}
\label{Eq:V_eff}
\begin{align}
    \beta V_{\text{ext}}(\mathbf{r}) &= [\ln(\rho(\textbf{r})) - \beta \mu]\alpha(\textbf{r}) \\
    \alpha(\textbf{r}) &= \sum_{i=1}^{M} \frac{\sigma^2}{r_c^{\gamma}},
\end{align}
\end{subequations}
where $\sigma$ is the in-sphere radius of a shape, $r_c$ is the distance to the surface of a shape added to the in-sphere radius (and thus $r_c$ is non-zero at the surface of a shape), and $\gamma$ must be larger than the spatial dimension of the system (we discuss the choice of $\gamma$ in a later section).
\noindent 
With an expression for $V_{\text{ext}}$ in hand, we substitute Eq.~\ref{Eq:V_eff} into Eq.~\ref{Eq:pP_F} (ignoring the constant energy shift $\ln[\frac{N^N}{N!}]$) to obtain: 
\begin{equation}
\label{Eq:Fe_eq}
\begin{split}
    F[\rho(\mathbf{r})]&= \\
    & \int d\mathbf{r} \big(\left(\alpha(\mathbf{r}) + 1\right) \rho(\mathbf{r})\ln[\rho(\mathbf{r})] - \alpha(\mathbf{r})\rho(\mathbf{r})\mu \big) \\
    &+ \int d\mathbf{r} \big(\lambda \rho(\mathbf{r})\big) -\lambda N.
\end{split}
\end{equation}
We can now easily solve for the equilibrium density field $\rho_{\text{eq}}(\textbf{r})$ by setting the functional derivative of $F[\rho(\textbf{r})]$ with respect to $\rho(\textbf{r})$ to zero. This procedure (detailed in the \textbf{SI}) gives us an expression for $\rho_{\text{eq}}(\textbf{r})$ and the minimized energy of the auxiliary system:
\begin{equation}
    \label{Eq:rho_eq}
    \ln \rho_{\text{eq}}(\textbf{r}) = \frac{\alpha(\textbf{r})(\mu-1)-\lambda-1}{1+\alpha(\textbf{r})}.
\end{equation}
\begin{equation}
    E_{\text{aux}} =  -\int d\textbf{r} \rho_{\text{eq}}(\textbf{r})\alpha(\textbf{r}).
\end{equation}
The energy $E_{\text{aux}}$ is the key quantity of interest for the auxiliary system. We assert that $-E_{\text{aux}}$ is analogous to the entropy of the original hard shape system, and that the arrangement of hard shapes that minimizes $E_{\text{aux}}$ will be identical to the arrangement of hard shapes that maximizes the entropy of the original system. In this way, we think of the quantity $\rho_{\text{eq}}(\textbf{r})\alpha(\textbf{r})$ as a spatially dependent local entropy density, whose integral over the entire auxiliary system gives us the entropy of the original system. We confirmed that the same expression for $E_{\text{aux}}$ is obtained via a self-consistent field theory approach based on the Hubbard–Stratonovich transformation~\cite{fredricksonFieldTheoreticSimulationsSoft2025}, assuming our hypothesized form of $V_{\text{ext}}$.

\textbf{\textit{Numerical Results --}}
We solved the EB-DFT equations for $\rho_{\text{eq}}(\textbf{r})$ and $E_{\text{aux}}$ to determine the thermodynamically preferred crystal structure for a variety of systems of hard particle shapes, initializing structures consistent with different candidate crystal macrostates in the auxiliary system by specifying the fixed positions and orientations of shapes in a discretized density field (details in the \textbf{SI}). We associate the auxiliary energy difference of two candidate structures with the entropy difference of the equivalent hard shape systems. The latter can be calculated for any stable or metastable colloidal crystal of hard shapes using the well-known Frenkel-Ladd (FL) free energy ($F_{\text{FL}}$) calculation method~\cite{frenkelNewMonteCarlo1984}, thereby providing ``ground truth'' values of the entropy against which to compare $E_{\text{aux}}$. In the case of hard spheres (the first example below), we can compare the EB-DFT predictions directly to the known equation of state~\cite{speedyPressureEntropyHardsphere1998,almarzaClusterAlgorithmMonte2009}.

In validating EB-DFT, choices arise as to what value to use for the exponent $\gamma$, which controls the steepness and range of $\alpha(\mathbf{r})$. We report results for $\gamma$ chosen to match reference free‑energy differences (Speedy EOS for spheres\cite{speedyPressureEntropyHardsphere1998,almarzaClusterAlgorithmMonte2009}; FL for anisotropic shapes) as well as for fixed $\gamma$ (we determined that the former route is most accurate). Simulations and FL calculations were performed using the hard particle Monte Carlo (HPMC)~\cite{andersonScalableMetropolisMonte2016} module of the open-source particle simulation toolkit HOOMD-Blue version 5.3.1~\cite{andersonHOOMDbluePythonPackage2020}, detailed in the \textbf{SI}.  

\textit{Validation of EB-DFT on the hard sphere system}\textbf{ --}
As an initial validation, we applied EB-DFT to the canonical hard sphere system and find the well-known result that the face-centered cubic (FCC) arrangement of spheres is just slightly more thermodynamically stable than hexagonal close-packed (HCP)~\cite{frenkelNewMonteCarlo1984}. To calculate $E_{\text{aux}}$, we chose $\gamma$ by fitting the EB-DFT energy difference between FCC and HCP to known values of the entropy difference from the Speedy equation of state~\cite{speedyPressureEntropyHardsphere1998}, with parameters given by Almarza~\cite{almarzaClusterAlgorithmMonte2009}, shown in Fig.~\ref{fig:sphere},  We also show how fixed values of $\gamma$ over- or underestimate the entropy difference, but still predict the stability of FCC over HCP at all relevant particle volume fractions $\nu$.
\begin{figure}[!htbp]
    \includegraphics*[scale = 0.5]{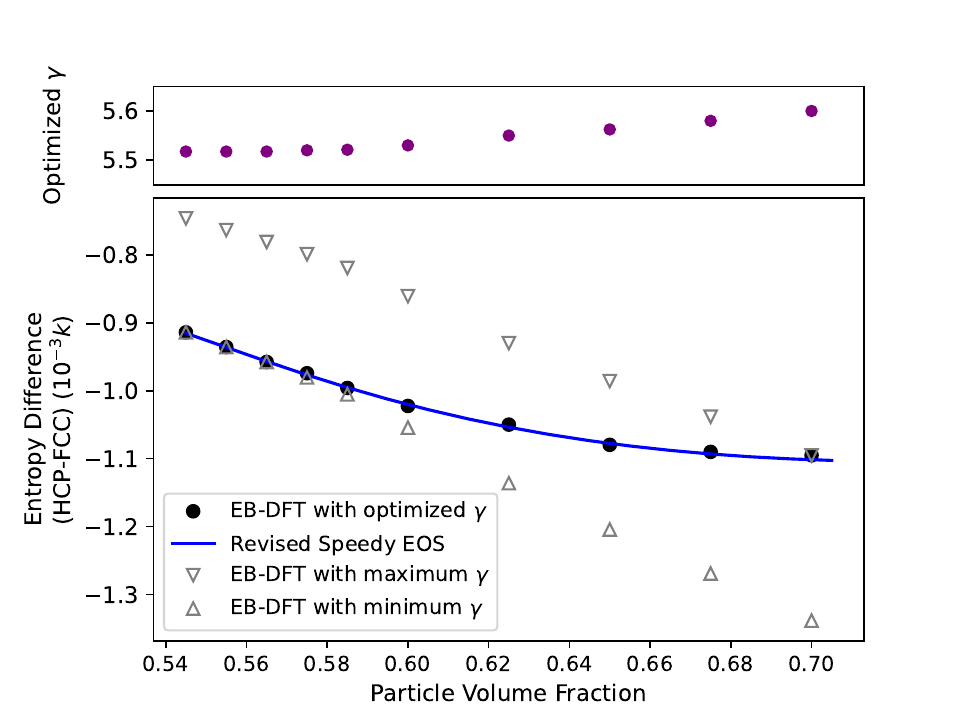}
    \caption{Entropy difference between HCP and FCC in a hard sphere system calculated from EB-DFT (black filled circles) fit, using the corresponding values of $\gamma$ in the upper panel, to the ``ground truth'' revised Speedy equation of state (blue curve) from~\cite{almarzaClusterAlgorithmMonte2009}.  The gray triangles show the EB-DFT prediction with fixed $\gamma$, using the smallest and largest values of $\gamma$ from the upper panel. Error bars, calculated as the standard deviation about the mean, are smaller than the symbols.}
    \label{fig:sphere}
\end{figure}

\begin{figure}[!htbp]
    \includegraphics*[scale = 0.5]{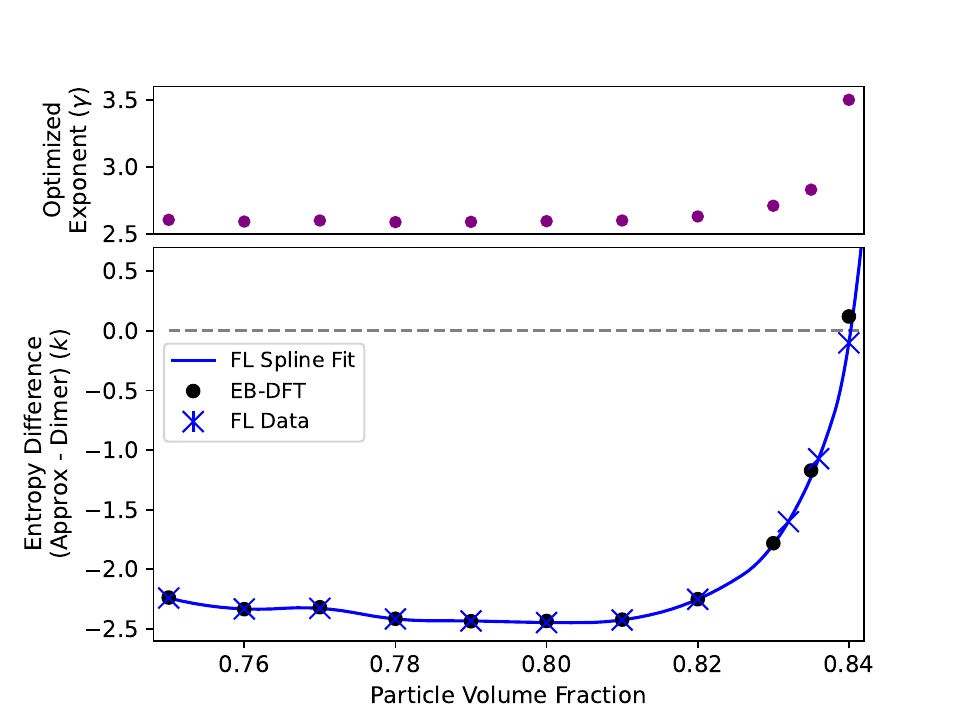}
     \caption{Entropy difference between dimer crystal and quasicrystal approximant in a hard tetrahedron system calculated using EB-DFT, fit to FL data along with the best fitting exponent $\gamma$ shown in the upper panel. Errors were calculated using the standard deviations of the values obtained from the simulation replicas, and are smaller than the symbols.}
    \label{fig:dimer}
\end{figure}

\textit{Validation of EB-DFT for a quasicrystal and a dimer crystal of hard tetrahedra}\textbf{ --}
From previous HPMC simulations and FL calculations, we know that hard, regular tetrahedra self-assemble into a dodecagonal quasicrystal at intermediate $\nu$~\cite{haji-akbariDisorderedQuasicrystallineCrystalline2009, wangControlledSelfAssemblyGold2023}, while the dimer crystal is thermodynamically preferred over the quasicrystal approximant at higher $\nu > 0.84$~\cite{haji-akbariPhaseDiagramHard2011,chenDensePackingRegular2008}. 
Fig. \ref{fig:dimer} shows  $\Delta E_{\text{aux}}$ calculated between the two crystals (solid black circles) using EB-DFT plotted vs.~$\nu$. The data changes sign at $\nu = 0.84$, in excellent agreement with the HPMC simulations.  We also fit $\Delta E_{\text{aux}}$ to $-\Delta S$ obtained with FL calculations (blue symbols), detailed in the SI, which gives us the $\nu$-dependent  exponent $\gamma$ for which EB-DFT best matches FL (shown in the inset). 

\begin{figure}[!htbp]
    \includegraphics*[scale = 0.5]{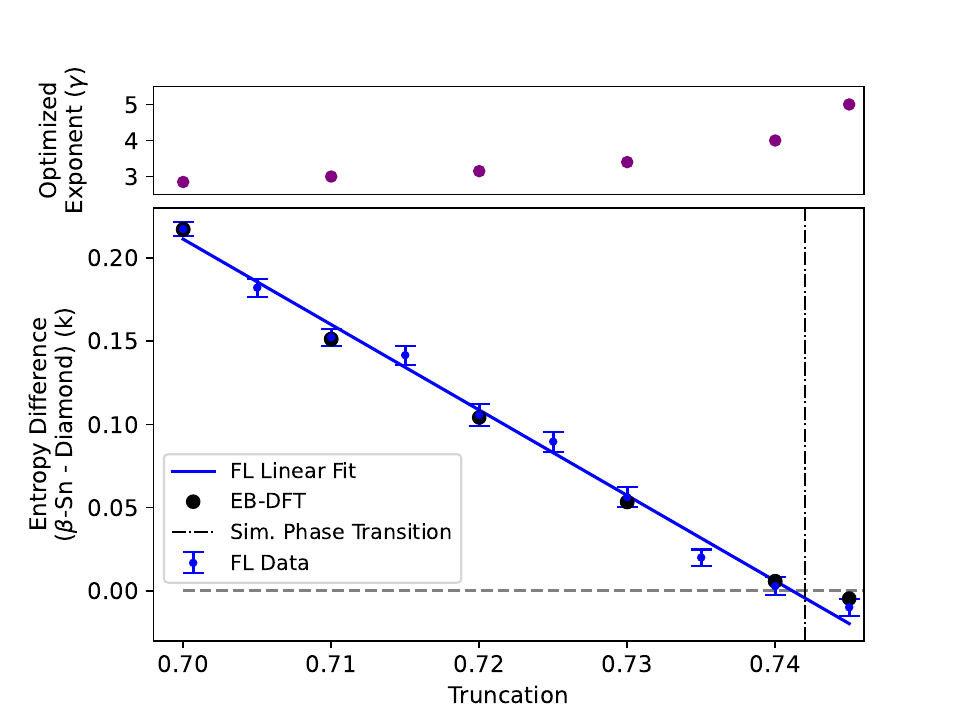}
    \caption{Entropy difference vs vertex truncation of a tetrahedron, where $t=0$ is a tetrahedron and $t=1$ is an octahedron. EB-DFT predicts the $\beta-Sn$ structure becomes more stable than diamond at a truncation value above $t\sim0.742$, in agreement with previous simulation results~\cite{damascenoCrystallineAssembliesDensest2012}. The values of $\gamma$ used in the calculations are shown in the upper panel. As in Fig. \ref{fig:sphere}, errors were calculated using the standard deviations of the free energies obtained from the simulation replicas.}
    \label{fig:trunctetra}
\end{figure}

\textit{Validation of EB-DFT for diamond and $\beta-Sn$ crystals of hard truncated tetrahedra --}
Having demonstrated that EB-DFT can correctly predict phase transitions as a function of particle volume fraction in the above two examples, we tested the ability of EB-DFT to predict phase transitions as a function of particle shape, at fixed volume fraction. As shown in Fig. \ref{fig:trunctetra},  EB-DFT correctly predicts the relative stability of diamond vs. $\mathbf{\mathit{\beta-Sn}}$ crystal phases in a system of hard, vertex-truncated tetrahedra, in agreement with previous HPMC simulations~\cite{damascenoCrystallineAssembliesDensest2012}. We observe that the value of $\gamma$ for which $\Delta E_{\text{aux}}$ matches $-\Delta S$ from FL calculations increases as the system approaches the phase transition, both as a function of density and shape. Understanding the interplay between shape and density will be key to improving upon the form of $V_{\text{ext}}$ in the future. We also note that the range of $\gamma$ for this example is larger than in the other examples, which is another indication that our simple expression for $V_{\text{ext}}$ should be improved in future work.

\textit{Validation of EB-DFT  for plastic (rotator) colloidal crystals of hard shapes—}
To further demonstrate our theory's ability to correctly predict crystal phases, we applied EB-DFT to seven different shapes that each form plastic (rotator) crystals~\cite{damascenoPredictiveSelfAssemblyPolyhedra2012,karasPhaseBehaviorDesign2019}. In the plastic phases, the positions of the particle centroids form a crystal, but the particles rotate about their centroids. Using $\gamma$ from our hard sphere example at the same volume fraction, EB‑DFT predicts the correct stable phase for all seven cases; details and the summary table are in the \textbf{SI}.

\textbf{\textit{Conclusion --}}
We presented a new formulation of classical density functional theory that correctly predicts equilibrium phases of hard particle crystals. Importantly, our cDFT formulation is fundamentally different from that typically used in liquid state theories of hard particle systems because we treat the hard particles not as solvent, but as fixed solute ``boundary conditions'' influencing a fictitious density field in an auxiliary system. This critical change of reference frame frees us from having to construct terms describing the interactions between hard particles (i.e., the ``excess'' potential), like those in White Bear and other fundamental measure theories~\cite{hansen-goosDensityFunctionalTheory2006}. Instead, our consideration of an auxiliary system in which a fictitious density field acts under an external potential produced by the hard shapes is more analogous to quantum DFT, where the electron density field in an auxiliary system of non-interacting electrons is subject to an external (Coulomb) potential produced by fixed, charged nuclei, which act as solute. Our motivation and justification for this new approach in which the hard particles are treated like ``atomic nuclei'' is the increasingly vast collection of colloidal crystals that are isostructural to atomic crystals.

We note that EB-DFT provides an alternative formulation of a theory of entropic bonding (EBT) developed by Vo and Glotzer by self-consistently solving an eigenvalue equation resulting from a convection-diffusion equation ~\cite{voTheoryEntropicBonding2022}. Using the alternate framework of classical DFT allows us to support the various assumptions and parameter choices in the theory with more intuitive physical underpinnings and fewer \textit{ad hoc} assumptions, and will, in future work, allow for direct comparison with quantum DFT calculations of atomic crystals. 

Of course, in quantum DFT of atomic systems, the electron-electron interaction (excess, or correlation-exchange term) combines with the external field to produce the spatially varying, equilibrium electron density distribution that corresponds to the arrangement of atomic nuclei in the thermodynamically preferred crystal structure. In our formulation of EB-DFT for hard shape systems, there is no exchange-correlation or excess term. The external potential, alone, is posited to produce the spatially dependent, equilibrium entropy density field corresponding to the arrangement of hard shapes in the thermodynamically preferred colloidal crystal structure. In improving upon the form of $V_{\text{eff}}$ in future iterations of EB-DFT, terms in addition to $V_{\text{ext}}$ may warrant consideration. Such consideration, or a refined expression for $V_{\text{ext}}$, may yield the benefit of eliminating $\gamma$ as a free parameter. We are also working to improve EB-DFT to accurately reproduce equations of state over a larger range of volume fractions and shape parameters, extending away from the relevant phase transition.

Finally, we note that a key feature of our theory is the mapping of entropy -- a global, system-level quantity -- to a local, spatially dependent ``entropy density field'' whose integration over the volume of the auxiliary system gives the entropy of the original system. A similar type of mapping has seen success in other contexts; e.g., the Oseen-Frank free energy~\cite{chaikinPrinciplesCondensedMatter1995}, which can describe hard particle liquid crystals with local variables. Our mapping gives us tools to describe, in a new way, the emergent statistical entropic forces that act locally to produce order in hard shape systems~\cite{harperEntropicBondColloidal2019,avendanoPackingEntropicPatchiness2017,vanandersUnderstandingShapeEntropy2014a}.

\textbf{\textit{Acknowledgements --}} This work was supported by a Vannevar Bush Faculty Fellowship sponsored by the Department of the Navy, Office of Naval Research under ONR award number N00014-22-1-2821. IRS was supported by the Department of Defense (DoD) through the National Defense Science and Engineering Graduate (NDSEG) Fellowship Program. This work used the Anvil system at Purdue University through allocation DMR 140129 from the Advanced Cyberinfrastructure Coordination Ecosystem: Services \& Support (ACCESS) program, which is supported by National Science Foundation grants 2138259, 2138286, 2138307, 2137603, and 2138296. The authors thank J.A. Anderson, T. Vo, D. Kofke and T. Waltmann for helpful discussions.

\bibliographystyle{apsrev4-1}
\bibliography{ref.bib}

\end{document}


\textbf{\textit{Supplementary Information --}}
\textit{Derivation of $E_{\text{aux}}$ --}
Taking the functional derivative of the free energy in the auxiliary system, we have:
\begin{equation}
    \begin{split}
        \left . \frac{\delta F[\rho(\textbf{r})]}{\delta \rho(\textbf{r})} \right |_{\rho=\rho_{\text{eq}}} &= \\
        & 1+\ln \left[ \rho_{\text{eq}}(\textbf{r}) \right] + \alpha(\textbf{r}) + \alpha(\textbf{r}) \ln[\rho_{\text{eq}}(\textbf{r})] \\
        &- \alpha(\textbf{r})\mu + \lambda \\
        &= 0.
    \end{split}  
\end{equation}
This minimization gives the following expression for $\rho_{\text{eq}}(\textbf{r})$:
\begin{equation}
    \label{Eq:rho_eq}
    \ln \rho_{\text{eq}}(\textbf{r}) =  \frac{-(\lambda+1)} {1+\alpha(\textbf{r})}
    + \frac{\alpha(\textbf{r})(\mu-1)}{1+\alpha(\textbf{r})}.
\end{equation}
\noindent We can now write the free energy by substituting Eq.~\ref{Eq:rho_eq} into Eq.~\ref{Eq:FE_eq},
\begin{equation}
    F[\rho_{\text{eq}}(\textbf{r})] = -\int d\textbf{r} \rho_{\text{eq}}(\textbf{r})\alpha(\textbf{r}) - N(\lambda+1)
\label{Eq:FE_eq}
\end{equation}
where the Lagrange multiplier $\lambda$ is found by normalizing the density:
\begin{equation}
    Ne^{\lambda + 1} = \int d\mathbf{r} (e^{\lambda + \mu})^{\alpha/(1+\alpha)}.
\label{norm_eq}
\end{equation}
To compare to the original hard shape system, we define the purely energetic part of this effective free energy:
\begin{equation}
    E_{\text{aux}} =  -\int d\textbf{r} \rho_{eq}(\textbf{r})\alpha(\textbf{r}).
\end{equation}

\textit{Computational Implementation --}
By exploiting symmetry, computation of the density field is restricted to within the unit cell of the given candidate crystal, even when scaling system size. For example, in a calculation involving eight unit cells, we compute the density field in only one of them, but use distances calculated from all eight unit cells. Specifically, increasing the number of shapes scales the distance computations linearly per sample point, but does not require additional sample points, enabling calculations on large systems. To eliminate finite-size concerns, all computations were performed on sufficiently large systems for which the results remain unchanged when the system size is halved.

Within the unit cell, we randomly sample the positions where we compute $\rho(\mathbf{r})$ with periodic boundary conditions. This allows us to estimate $E_{\text{aux}}$ via Monte Carlo integration~\cite{metropolisMonteCarloMethod1949}, which also allows us to to estimate the error on $E_{\text{aux}}$. For every computation, we use a sufficient number of samples so that the error in $E_{\text{aux}}$ is significantly smaller than the magnitude of $E_{\text{aux}}$, to the point where the error bars are no longer visible in the plots. We developed and implemented a new algorithm for computing minimum distances between shapes and points, which enabled us to use a large number of samples for our Monte Carlo integration. This algorithm will be released in a future version of the Python library coxeter~\cite{Ramasubramani2021} (https://github.com/glotzerlab/coxeter).

The normalization factor ($\lambda$ in Eq. \ref{Eq:FE_eq}) is found using Newton's method to solve Eq. \ref{norm_eq}, where the log of the solution is used to avoid issues due to regions where the integrand is large. We solved for $E_{\text{aux}}$ and $\rho_{\text{eq}}(\textbf{r})$ for incrementally increasing values of $\mu$ to determine the point at which $E_{\text{aux}}$ no longer varied significantly with $\mu$. At these large $\mu$ values, the auxiliary system can be considered ``saturated'' by the density field; that is, in the limit of large $\mu$, the density field is incompressible. We note that the normalization of the density prevents the total energy from diverging.
 
\textit{Entropy Calculations --}
Entropy calculations were performed using the Frenkel-Ladd method \cite{frenkelNewMonteCarlo1984}. Since our systems consist of anisotropic particles with rotational degrees of freedom, we included additional springs to tether particles to their average orientations in the lattice. In simulation, the spring constant for the translational and rotational springs, $k$ $\in$ [$0$,$10^{10}$], was gradually weakened starting from the largest value. At each spring constant increment, the simulations ran for $2\times10^5$ time steps, and average energies of the particles were calculated throughout. To avoid drift, the center of mass of the system was re-centered at each timestep. Simulations of tetrahedra and truncated tetrahedra were replicated 5 and 10 times, respectively, for statistics, and the corresponding entropy differences were calculated via integration of the averaged energies across spring constants. The standard error was computed from the standard deviation of the free energies calculated from the independent simulation replicas.

\begin{table*}[h]
\caption{EB-DFT predictions of plastic crystal (rotator phase) stability for several shapes. For each shape, EB-DFT correctly identifies the stable phase reported in prior literature. ($\phi$ = volume fraction, p prefix represents a plastic phase). The original classification of the plastic crystals of the Truncated Cuboctahedron, Rhombic Dodecahedron, and the Pseudorhombicuboctahedron come from Karas, et al~\cite{karasPhaseBehaviorDesign2019}, while the Triakis Octahedron, Trucated Octahedron, and Cuboctahedron are from Damasceno, et al~\cite{damascenoCrystallineAssembliesDensest2012}.}
\begin{ruledtabular}
\begin{tabular}{lccccc}
Shape & Structure 1 & Structure 2 & $\phi$ & EB-DFT prediction & Simulation prediction\\
\hline
\hline
Truncated Cuboctahedron & pFCC & BCT & 0.54 & pFCC stable  & pFCC stable \\
Rhombic Dodecahedron & pFCC & FCC & 0.54 & pFCC stable  & pFCC stable \\
Rhombicuboctahedron & pFCC & FCC & 0.54 & pFCC stable  & pFCC stable \\
Pseudorhombicuboctahedron & pFCC & BCT & 0.54 & pFCC stable  & pFCC stable \\
Triakis Octahedron & pBCC & FCC & 0.54 & pBCC stable  & pBCC stable \\
Truncated Octahedron & pBCC & BCC & 0.54 & pBCC stable  & pBCC stable \\
Cuboctahedron & pFCC & BCC & 0.54 & pFCC stable  & pFCC stable \\
\end{tabular}
\end{ruledtabular}
\label{tab:plastics}
\end{table*}
\textit{Plastic Crystal Simulations --}
Hard particle Monte Carlo simulations were performed using the HOOMD-blue package version 5.3.1~\cite{andersonHOOMDbluePythonPackage2020}. As simulation data for the spheres, tetrahedra and truncated tetrahedra were taken from the literature, the only new MC simulations performed here were for the plastic crystals. Previous publications reported plastic crystal phases and their shapes, but additional information was required for our purpose.  For each shape, we compared the stability of two candidate structures at fixed volume fraction, and  evaluated which has the lower value of $E_{\text{aux}}$ without parameter fitting (we chose $\gamma$ as for spheres of the same volume fraction).  To correctly sample orientations in the plastic crystals, we prepared the systems by running HPMC simulations at volume fractions above and below the phase transition. The higher density phase was expanded to the same density as the lower density phase, and we then computed the difference in $E_{\text{aux}}$ between the two phases and compared the EB-DFT predictions to known results from prior HPMC simulations~\cite{karasPhaseBehaviorDesign2019,damascenoCrystallineAssembliesDensest2012}. Plastic crystal systems were initialized in cubic simulation cells with periodic boundary conditions containing 512 particles. Each system was equilibrated for $100\times10^6$ time steps using move sizes calibrated to achieve $20\%$ acceptance rates. We additionally incorporated isochoric moves that could shear or modify the simulation cell aspect ratio to relax strain imposed by simulation cell symmetry constraints using the BoxMC HOOMD-Blue module in version 5.3.1~\cite{andersonHOOMDbluePythonPackage2020, andersonScalableMetropolisMonte2016}. Crystal structures were determined via Common Neighbor Analysis in the OVITO software package~\cite{fakenSystematicAnalysisLocal1994,honeycuttMolecularDynamicsStudy1987,ovito}.

\bibliography{ref}